# Role of Helium Bubbles and Voids on the Property of Nanocrystalline Tungsten

Yang Zhang[1] and Jason R. Trelewicz[1,2,*]

[1]Department of Materials Science and Chemical Engineering, Stony Brook University, Stony Brook, NY 11794

[2]Institute for Advanced Computational Science, Stony Brook University, Stony Brook, NY 11794

*Corresponding Author at Stony Brook University: jason.trelewicz@stonybrook.edu

**Abstract:**

This manuscript embarks on an inquiry into the influence of helium implantation on nanocrystalline tungsten, a contender for plasma-facing components (PFCs) in nuclear fusion reactors. The study underscores the inevitability of helium retention in tungsten due to the anticipated high flux of helium atoms from D-T fusion reactions in future reactors such as ITER and DEMO. This retention, potentially culminating in surface blistering and grain boundary embrittlement, propels an investigation into the helium's preferential substitutional configuration over interstitial within tungsten's lattice, leading to cavity and bubble formation. Atomistic tensile testing simulations, exploiting a helium-specific interatomic potential, dissect the ramifications of helium concentration and voids on tungsten's mechanical properties across grain sizes of 10nm and 15nm. The presence of helium at grain boundaries instigates detachment, influencing structural behaviors under strain. Notably, the study reveals a dichotomy in helium's effect: benign at concentrations up to the critical point, beyond which significant embrittlement and elastic softening occur, corroborating with theoretical predictions on helium-induced grain boundary embrittlement. This nuanced exploration delineates the intricate dance between helium implantation levels, grain size, and the ensuing mechanical property alterations in nanocrystalline tungsten, casting light on its viability and endurance as a PFC material under the rigorous conditions forecasted in fusion reactors.

## 1.1. Introduction

As one of the candidate materials for PFCs, the nanocrystalline tungsten will experience a significant amount of helium implantation as a product of the fusion reaction. For example, in ITER and DEMO, the D-T fusion reaction is estimated to happen at a rate above $10^{21}$ s$^{-1}$ [1, 2], resulting in a helium flux above $10^{18}$ m$^{-2}$s$^{-1}$ on the first wall materials. For a single 400s long pulse operation, the fluence of the helium implanted in the FW materials can reach $10^{20}$ m$^{-2}$. During a full-power-year operation, $10^{23}$ m$^{-2}$ helium fluence can be easily achieved. Gilliam et al. [3] have shown that at this level of helium fluence (and fluence per implant-anneal cycle), the helium implanted in the tungsten may not be removed by the heat load cycles during operation. Therefore, helium retention in the tungsten may play a critical role in predicting the behaviors of the material. Their results also indicated that the peak helium concentration in the tungsten above 4 at% will result in the surface blistering. Surface blistering has been proposed to be a result of plastic deformation by the formation of cavities and bubbles in the sub-surface or near-surface layer [4]. The mechanism of He bubble formation has been proposed to be a result of the extremely low solubility in tungsten: the helium atom prefers to be in a substitutional configuration than in the interstitial configuration in tungsten [5]. More and more observations demonstrated the formation of cavities and gas bubbles in the tungsten [6-11] due to helium retention. The helium segregation to the grain boundaries may destabilize the grain structure when the stored energy associated with He at grain boundaries becomes close to the energy required to turn grain boundaries into free surfaces ($E_{He}^{sol} v_{He}^c \approx 2\varepsilon_{surf}$, where $E_{He}^{sol}$ is the energy of solution for He at substitutional site, $v_{He}^c$ is the He critical boundary density, and $\varepsilon_{surf}$ is the surface energy) [12, 13]. Therefore, the He-induced grain boundary embrittlement occurs when grain boundary He reaches the critical boundary density, which corresponds to a critical bulk concentration ($C_{He}^c = 3 v_{He}^c / an$, where a is the average grain size, and n is atmic density of the material). Based on this equation, the grain size effect shall be pronounced on the He-induced grain boundary embrittlement.

In the present work, atomistic tensile testing simulations were performed on ten structures: five with 10nm grain size, and the five with 15nm grain size. The grain structures in each grain size group are identical. By varying the helium concentration and the chemical environment, the helium effects on grain boundary embrittlement and their microscopic behaviors associated with structural deformation were revealed. The grain detachment due to the presence of the helium atoms at grain boundaries was demonstrated. Besides the grain boundary embrittlement effect, the comparison of the results also indicated the strength softening effect if the helium solutes at grain boundaries.

### 1.2. Simulation Methods

The interatomic potential selected for this study was specifically designed for the helium bubbles in tungsten [14], and most of the analyzing and visualization artwork was done with OVITO software [15].

#### 1.2.1. Sample Preparation and Tensile Testing Setting

Two nanocrystalline tungsten structures with an average of 10nm grain size and 15nm grain size were fabricated as the base structure. Each structure contained 25 grains, and the grain size distributions were optimized by Mote Carlo procedures, as shown in Figure 0-1. The total sizes of the structures are $(23.6 \text{ nm})^3$ and $(35.5 \text{ nm})^3$, respectively.

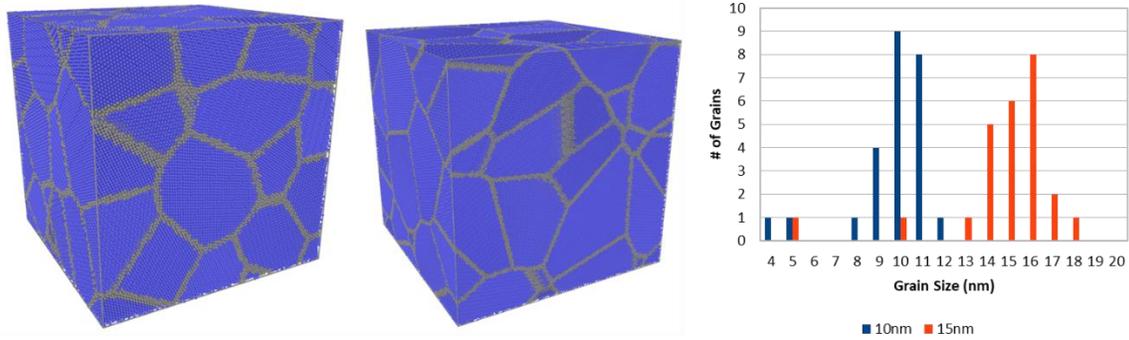

Figure 0-1 Nanocrystalline Tungsten structures with 10nm grain size and 15nm grain size. The right plot shows the grain size distribution in two structures.

Two sets of W-He structures were built with 1 at% of He and 5 at% of He, respectively. Based on the Ref. [13], the critical bulk concentration for grain boundary embrittlement in tungsten with 10nm grain size is 4.4 at%, and 2.9 at% for the structure with 15nm grain size. The concentrations selected for the present work may show the behavior below the critical concentration and above the critical concentration. The addition of He atoms was according to an MC-MC combined routine [16], allowing gradual replacement of tungsten atoms from most favorable sites for helium implantation. The structure was optimized to the minimal energy configuration during the 100 MC steps seeded in 100000 steps of MD relaxation. Then in the replicas of W-He structures, the He atoms were removed to create two structures with void decorated grain boundaries. All ten structures were then relaxed at 300K for 1ns to release internal stress.

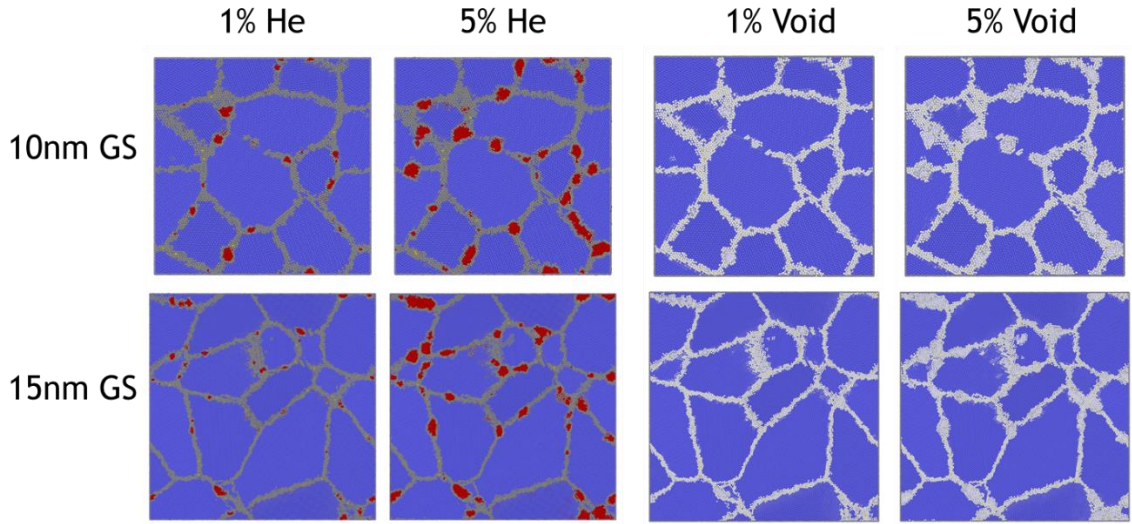

Figure 0-2 Pristine structures for uniaxial tensile testing with different grain sizes and different compositions of He and void. The blue dots are BCC tungsten atoms, the grey dots indicate the grain boundary atoms, and the red dots are helium solutes.

Since the structures within each grain size group were identical, the direct comparison within each grain size group excluded the structural effects, such as the grain boundary configuration and grain size distribution, which made the analyzing process more reliable.

The uniaxial tensile testings were performed on the LAMMPS platform [17], at a fixed strain rate of $10^7$ s$^{-1}$. The strain rate was selected based on the results from trial simulations at different strain rates, as shown in Figure 0-3. The trial simulations demonstrated that the slip behaviors could be sampled at a $10^7$ s$^{-1}$ strain rate. Though it would save simulation time at higher strain rates, the slip behavior could not be exhibit. Further reduction of the strain rate would provide more reliable results, but time consumption would make it costly.

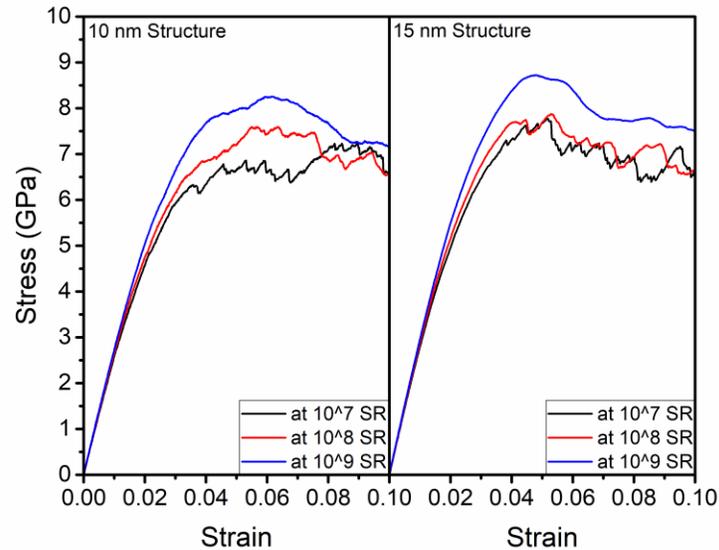

Figure 0-3 The stress-strain curve of uniaxial tensile testings at different strain rates up to 10% strain. The smooth curve at $10^9$ s$^{-1}$ strain rate showed no slip activity, while the curve at $10^7$ s$^{-1}$ strain rate indicated a significantly higher amount of slip activities.

### 1.2.2. Slip Activity Characterization

As the slip behavior was related in this study, but no specific algorithms were programed for the slip in BCC materials, the author developed one specifically for tungsten. The calculation was optimized by the lattice parameter of the BCC Tungsten and would reduce the noise significantly. First, the dislocation was characterized by the atoms in the slipped planes. The slip vector was used to pick out the slipped atoms, exclude the atoms in the GBs. Theoretically, the rest should be roughly the atoms in the slipped planes. However, in some regions of the lattice, the BCC atoms had random pointed slip vectors, which could be a result of grain rotation or local recrystallization during deformation. As those lattice atoms could not be classified correctly, modification of the algorithm was in need.

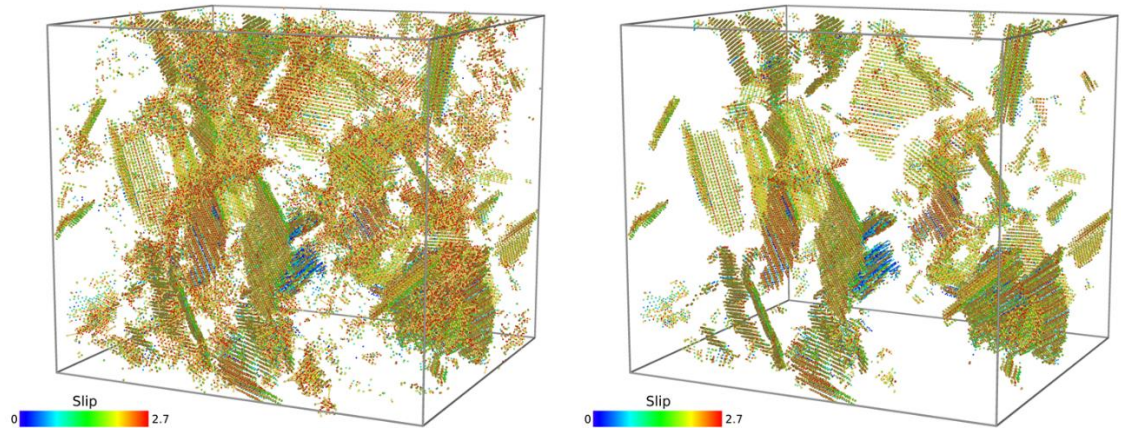

Figure 0-4 The slipped planes in the structures before and after the deviation-based noise reduction algorithm.

The method used to exclude those atoms was based on the deviation of the slip vector in the first nearest neighbor. In the slipped planes, the atoms were slipped to the same direction at a similar distance, the deviation in the lattice was remarkably lower than that of the atoms randomly displaced. This simple method provided significant improvement in the accuracy of the classification, the structural comparison before the noise reduction, and after the noise reduction is shown in Figure 0-4.

### 1.3. Structural Differences between He Implanted and Void Decorated Tungsten

The effect of the He and Void on structural properties were significant. Due to the considerable internal pressure in the He bubbles, the implantation of He reduced the average atomic volume of the lattice atoms. In contrast, the existence of voids increased the average atomic volume of the lattice atoms, as shown in the right figure of Figure 0-5.

The pressure in the He bubble was measured to be in the GPa range and showed a composition dependency and grain size dependency as in the left figure of Figure 0-5. The He bubble pressure differences in different structures didn't reflect on the atomic volumes of the lattice atoms, and the reduction in atomic volume relative to the pure structure is insignificant. However, the W-void structure demonstrated a significant change, probably due to the stress reduction by the free volume in the structure.

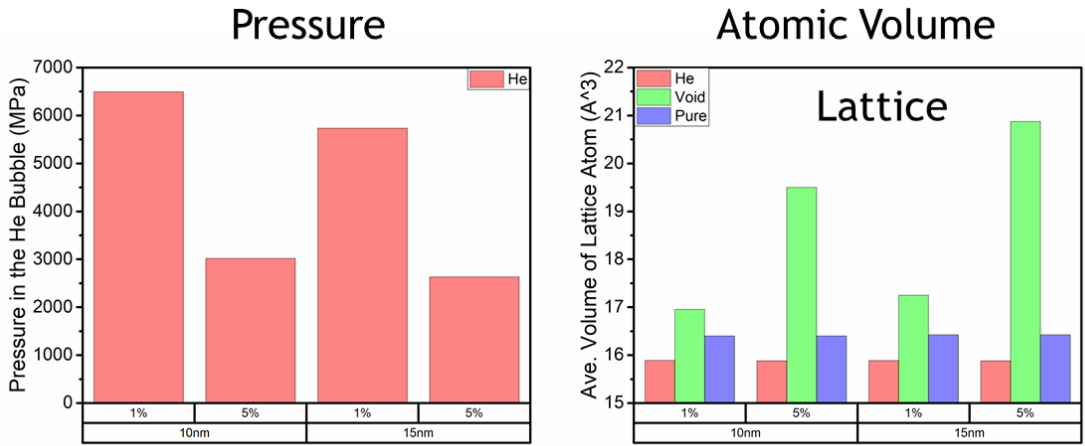

Figure 0-5 The pressure in the He bubble and the average atomic volume of the lattice atoms.

Based on the information of the bubble pressure and the atomic volume, one may assume that the W-He structure would show more internal stresses. However, the atomic stress maps showed an opposite picture. As shown in Figure 0-6, the He atoms in the GB helped in removing stresses in the GB and created a more homogenous stress distribution in GB. The possible answer to this phenomenon is that the substitutional He atoms may reduce the grain boundary density, as the He atoms are much smaller than the tungsten atoms.

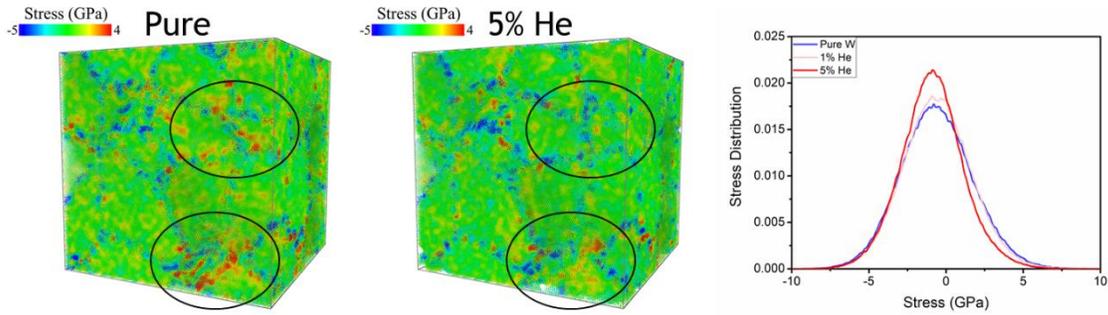

Figure 0-6 The atomic stress distribution in 15nm GS structures of pure tungsten and tungsten with 5at% He. The shape of the atomic stress distribution in W 5at% He is sharper near 0 pressure.

The atomic stress vector image of the 15nm W 5at% He structure even indicated some grain detachment, as shown in Figure 0-7. The stresses at the GB of the 15nm W 5at% He structure were less likely pointing perpendicular inward the grain, suggesting that two grains on the opposite sides of the grain boundary were not pressing each other. The lack of interaction between the grains may serve as a sign of detachment.

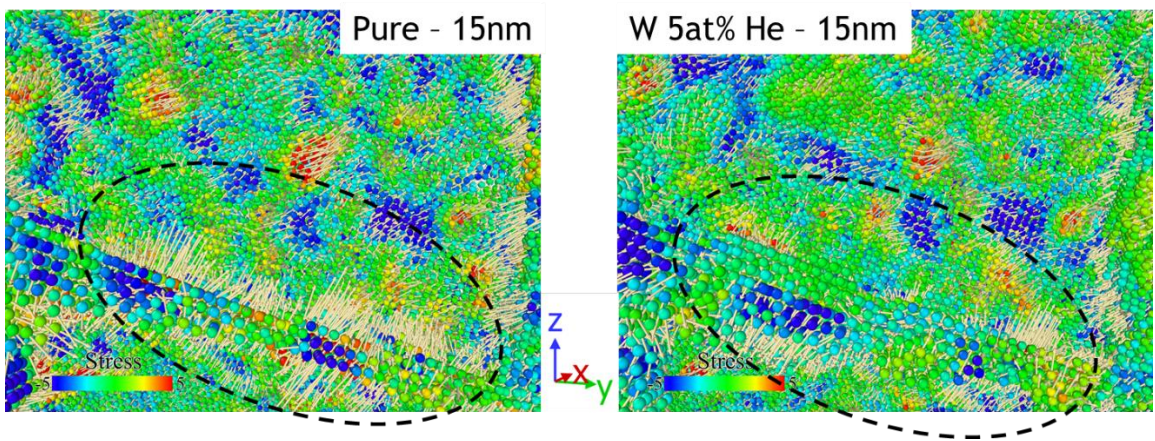

Figure 0-7 Atomic stress map in the 15nm grain size structures with different He concentration. The atomic stress in W 5at% He structure shows that the stress pointing from each side of GB to the other side is not as strong as in the pure structure, which suggests grain detachment.

The evidence of detachment was also found in the section view of the He implanted structure, as shown in Figure 0-8. Instead of in the round shape bubbles, some of the implanted He atoms formed into flat shape bubbles at the grain boundaries. Those flat shape bubbles with considerable pressure inside may cause the grain detachment in the material due to the weak interaction between He and W. For smaller grain size or smaller concentration of He, the GB may maintain the integrity and apply enough tension on the bubbles, thus reduced the grain detachment effect.

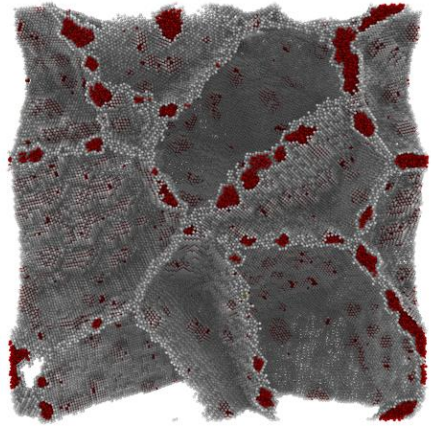

Figure 0-8 Section view of the W 5at% He with 15nm grain size. The grey dots are grain boundary atoms, and the red dots are He atoms. Some of the flat-shaped He bubbles break the integrity of the grain boundary, separate the grains on the opposite sides of the grain boundary, and cause the grain detachment.

### 1.4. Elastic Softening with 5 at% of He and Void

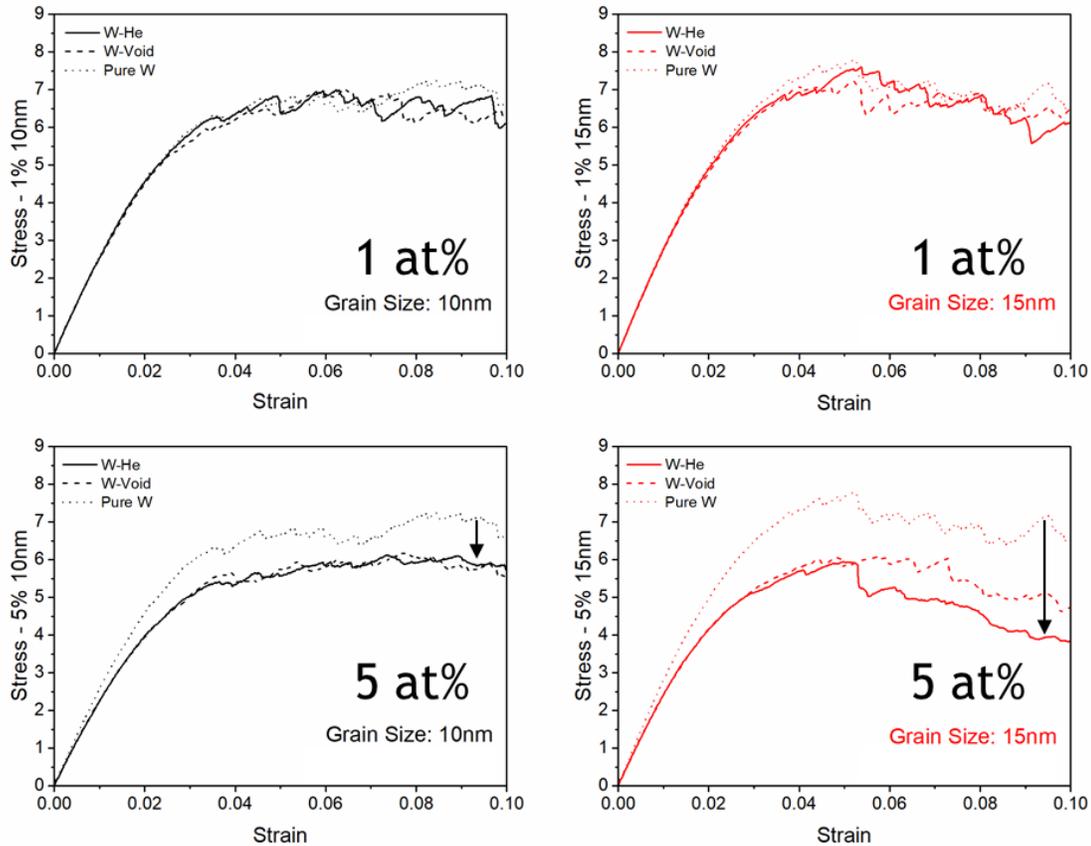

Figure 0-9 Stress-strain curve of each structure, grouped by the grain size and composition.

As shown in Figure 0-9, the stress-strain curve was recorded during the simulation, discernible softening in the structures with 5 at% He or Voids. The Young's Modulus was calculated at 0.1% strain, and the values were shown in Figure 0-10. The results showed a 10% reduction in the structure with 5at% He/Void. In contrast, the mechanical strength in the structures with 1at% He/Void didn't show significant degradation. To sum up, the implantation of He and Void showed a consistent effect on elastic properties. The cavities, whether filled or empty, may reduce the elastic strength of the structure in the same manner. It is explainable from the understanding of the elastic behavior related to the strong elastic interactions between atoms in the material, and the presence of the cavities may break the continuity of interaction transmission, in turn, reduces the elastic strength.

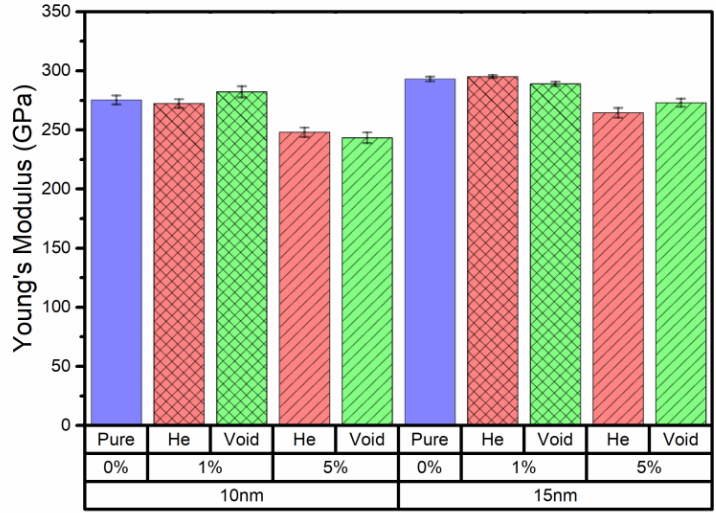

Figure 0-10 Young's Modulus of each structure. The tungsten structures with 1at% He or void do not exhibit degradation, while the structures with 5at% He or void show a 10% reduction in Young's Modulus.

The concentration dependent softening effect has been discussed in previous works as effective medium theories. Since the "pore" like helium bubble or void has negligible elastic modulus, the structure containing homogeneously distributed pores will show a reduced elastic modulus [18-20], based on the equation of: $E_r = \frac{E}{E_0} = (1-\phi)^2$ [18], where E is the elastic modulus of the structure with void/He bubble arrays, $E_0$ is the elastic modulus of the perfect structure, and ϕ is the void fraction (porosity). Weerasinghe et al. [21] performed tensile testing MD simulations on the single crystal tungsten with the nano-sized void or helium bubble arrays. Although the helium bubbles introduced significantly high stress in the structure, identical trends of Young's modulus and the shear modulus as a function of the void fraction (porosity) were revealed despite the composition of the void. This behavior indicates that the basic mechanisms of the void and helium effect on those moduli are identical under the settings of these MD simulations. A more recent MD study by Chen et al. [22] reaffirmed the softening effect of the dissolved He in the tungsten lattice. In experiments, the softening behavior was also observed by Liu et al. [23] in their micro-tensile testings on α-Zr. The Zr samples with different bubble size distributions showed that the deformation mechanism changed from the bubble-dislocation interaction induced hardening (bubble diameter less than 8 nm) to the bubble coalescence dominate plastic deformation (bubble diameter higher than 8nm). Our simulations were not set exactly as any of the above, but the bubbles and the voids were distributed at GBs exhibiting maily the softening effect, as suggested in the effective medium theories. The hardening by the dislocation-bubble interaction was suppressed, shown in the following sections.

### 1.5. Helium-Induced Grain Boundary Embrittlement

As shown in Figure 0-9, the 15nm W 5 at% He structure showed a sharp drop of strength after 5% strain, indicating the grain boundary embrittlement, which may be a result of the grain detachment. The plastic deformation was revealed based on the slip behaviors in the structure, as shown in Figure 0-11. When external strain applied, the plastic deformation differed significantly between W 5at% void and W 5at% He structures. Fewer slip activities were observed in W 5at% He structures, indicating that the activities other than slip accommodated a considerable amount of the plastic deformation. As indicated in the previous section, the implantation of the He reduced the interaction between the grains. The separation of the grains may cause a significantly lower plasticity of the structure.

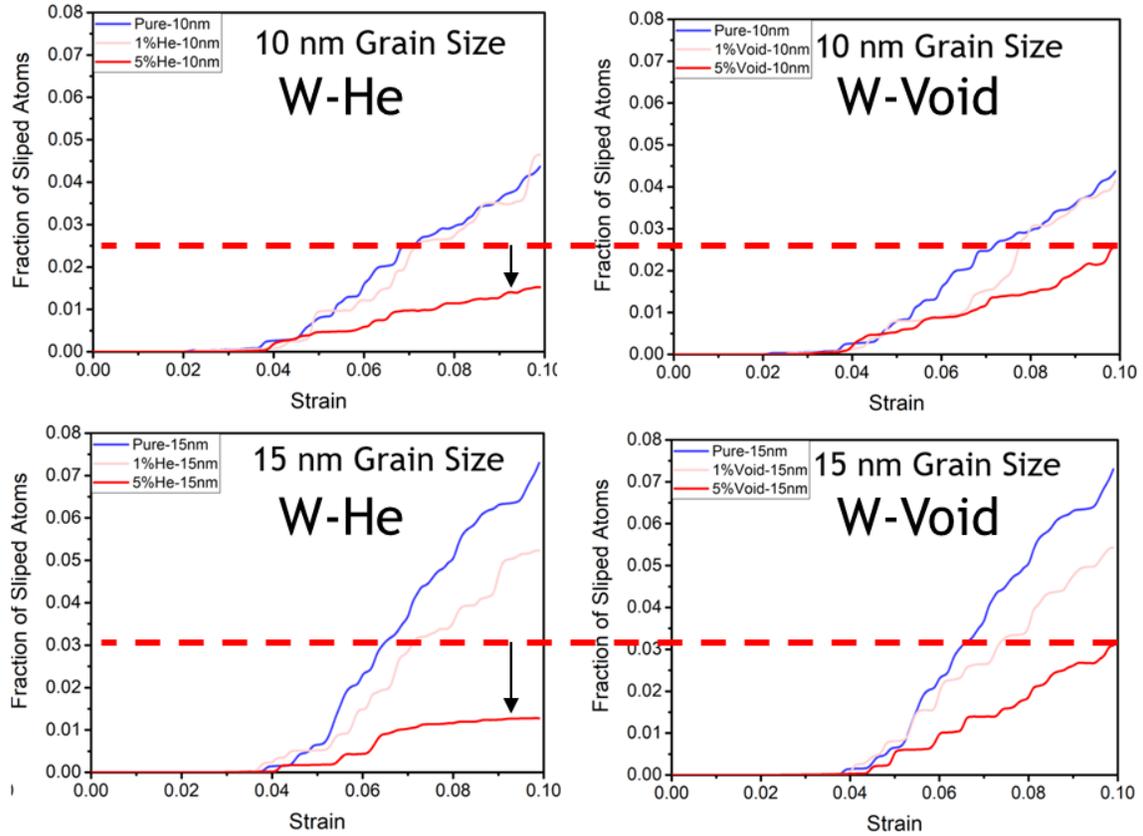

Figure 0-11 Fraction of Slipped atoms in W-He and W-void structures. The structures 1at% He or void show identical slip behaviors. while the

As shown in left plot of Figure 0-12, the cavity expansion and growth behavior were captured by the porosity of the structure, which was defined as the total volume occupied by He or Void. As discussed in the previous section, the pressure of the He bubble was extremely high. During the tensile testing simulation, the increase of the porosity showed an expansion of 40% for the bubbles in the 15nm grain size W 5at% He structure, while the corresponding W 5at% void structure showed no significant porosity change. The comparison of the W-He and W-void results indicated that the expansion and growth of the bubbles at grain boundaries accommodated the rest of the plastic deformation during tensile testing.

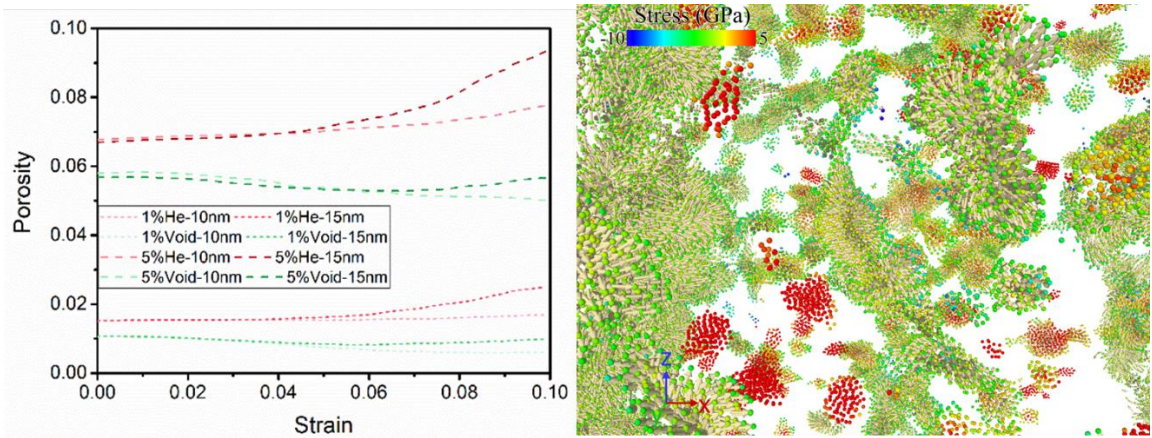

Figure 0-12 Left: The occupancy fraction of the bubbles and voids in the structures during tensile testing. The red lines represent W-He structures, and the green lines represent the W-void structures. The 15nm grain size structures are with darker colors, and the 10nm grain size ones are with lighter colors. Right: The growth of the bubble during the deformation. The vectors indicate the directional displacements of the He atoms, and the color indicates the atomic stress.

The mean-squared displacement (MSD) was calculated to validate this conclusion, as shown in Figure 0-13. The MSD proved the integrity between GB atoms and lattice atoms since these two components showed similar displacement behavior. The comparison between different compositions also revealed the He softening effect: with a higher concentration of helium, the tungsten atoms became twice as mobile as in the pure structure, indicating that the presence of He atoms at the GBs promoted the motion of the whole grain. The MSD of the He atoms in different structures revealed the significant bubble coalescence that happened in 15nm W 5at% He structure: most of the He atoms moved to adjacent bubbles at the end of the test given an MSD value around 1000, indicating an average travel distance of 3.2 nm. The coalescence of the bubbles intensified the grain detachment, which caused the He-induced grain boundary embrittlement.

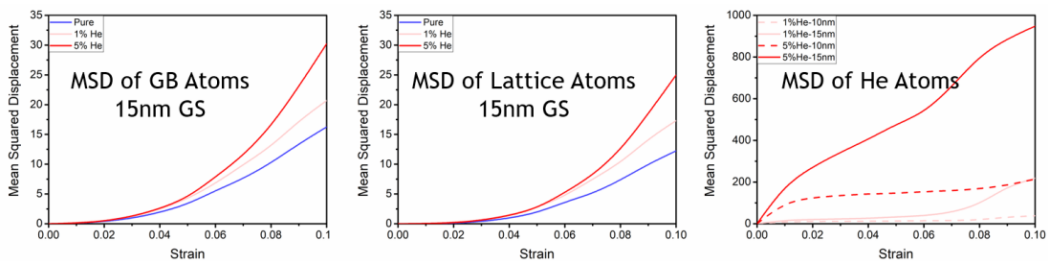

Figure 0-13 The mean squared displacement of different components in 15nm structures with various He concentration. The left two are the MSD of tungsten atoms, and the right one are the MSD of He atoms. Noting that the scale of the He MSD is 30 times larger than the tungsten MSD.

The expansion was also observed directly by the snapshots at different strains. The Helium bubbles didn't grow to round bubbles but grew to flat bubbles filling in the gaps between grains, which detached the grains. Therefore, He implantation demonstrated a grain boundary embrittlement effect. Besides, the detachment also reduced the stresses applied to the grains so that the slip activities were significantly suppressed in the W 5at% He structures.

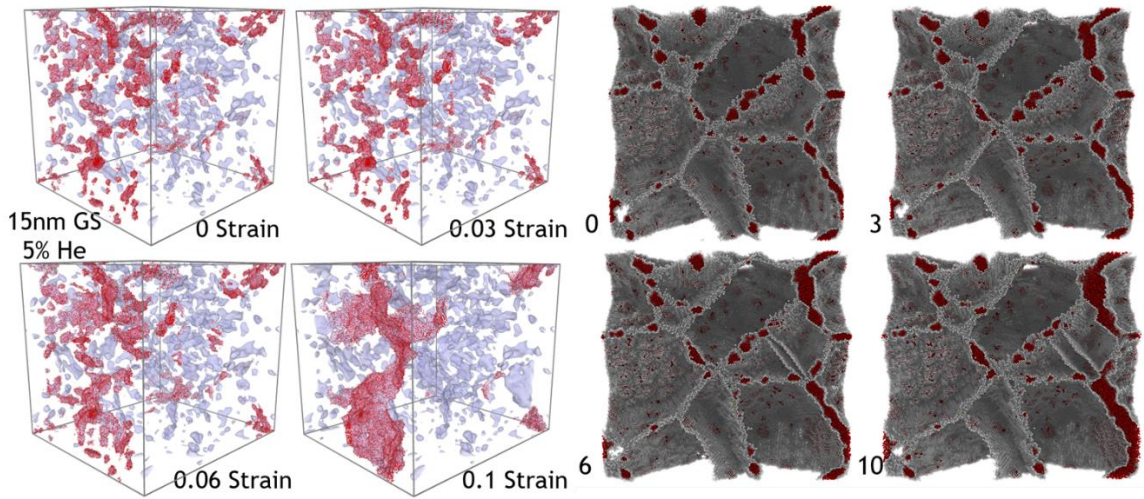

Figure 0-14 Left: The growth of the He bubbles during tensile testing. The red dots are the He atoms forming the largest bubble at the final strain, and the grey bubbles are the rest of the He bubbles in the structure. Right: The intensified detachment of grains during tensile testing. The index of the snapshot shows the percentage of strain in the snapshot. Red atoms are He atoms, the grey atoms are tungsten GB atoms.

The GB embrittlement behavior was reported in the aforementioned work by Chen et al. [22]. The tensile testing simulations on the structures containing helium clusters at GB demonstrated the GB embrittlement by the helium cluster growth at the GB during the tensile deformation. The bubble growth at the GB accommodated most of the plastic deformation, and due to the heterogeneous growth along the GB plane, the GB embrittlement was presented. The microstructure evolution indicated that the creation of the new surfaces at GBs leads to the embrittlement at relatively low strain. This behavior was predicted in a numerical model developed by Gilbert et al. [13] with a general idea that when the He atoms at the grain boundaries split apart the material by creating new surfaces, the failure due to GB embrittlement will take place. The critical bulk He concentration ($G_{He}^c$) proposed by Gilbert et al. was a function of average grain size (a), grain boundary energy ($\varepsilon_{surf}$), solution energy of the He at a substitutional site in a perfect lattice ($E_{He}^{sol}$), and the atomic density of the bulk material (n), as $G_{He}^c \approx \frac{6 \cdot \varepsilon_{surf}}{a \cdot n \cdot E_{He}^{sol}}$. This simplified model was used to predict the He-embrittlement-limited component lifetimes of the PFM under the worst-case scenarios. The authors stated that the estimation was based on the worst case scenario which might not fit for every case.

Based on this theory, for a tungsten material with an average grain size of 10 nm, the critical concentration of helium to induce the GB embrittlement was ~4.3%, and the value for a 15 nm grain tungsten was ~2.9%. Therefore, a material with an average grain size of 10 nm can withstand more helium implantations than the 15 nm grain material, in the term of helium induced GB embrittlement. Our simulations gave a good interpretation of the predicted behavior in the material under tensile strain.

### 1.6. Conclusion

Based on the mechanical property results and the morphology images, despite the difference between the Helium atoms and voids, their effects on the elastic behaviors of the structures were similar. Softening on elastic strength were observed in both structures with 5at% He or 5at% void, while the insignificant change was showed in W 1at% He and W 1at% Void structures. The elastic strength softening may relate to the discontinuity of grain interactions, regardless of the source.

The critical amount of discontinuity of the grain boundary is required to exhibit the discernible softening effect. A similar conclusion has been reported recently [21] with the cavity inside the single crystal tungsten.

However, significantly different effects of He atoms and voids on plastic behaviors were observed. The 15nm W 5at% He structures showed a significant grain boundary embrittlement effect. The behaviors of He can be summarized:

1. The presence of He atoms in the GB caused grain detachment, which reduced slip activities.
2. With 1at% of He or Void in the structure, the detachment effect was limited.
3. The inner pressure of the He bubbles was concentration-dependent as well as grain size-dependent. Lower concentration and the smaller grain size resulted in higher inner pressure. However, small bubbles will also experience high tension from the tungsten interfaces.
4. The inner pressure drove the bubbles to expand and coalesce. The result of this behavior intensified grain detachment, which caused the grain boundary embrittlement, and easier to dismantle.

The results from this work imply that up to 1at% of He implantations may not cause significant mechanical property degradation, which is consistent with previous prediction [13]. And the smaller grain size of the base material will help reduce the degradation effect caused by He. Further works need to be done to confirm the conclusions as the present work only took tensile strength into account.